\documentclass[twocolumn,amsmath,amssymb,pra]{revtex4}
\usepackage{graphicx}
\usepackage{dcolumn}
\usepackage{bm}

\begin{document}

\title{High-fidelity continuous-variable quantum teleportation\\toward multi-step quantum operations}

\author
{Mitsuyoshi Yukawa$^{1,2}$, Hugo Benichi$^{1,3}$, and Akira Furusawa$^{1,2}$}

\affiliation{$^{1}$Department of Applied Physics, School of Engineering, The University of Tokyo,\\ 7-3-1 Hongo, Bunkyo-ku, Tokyo 113-8656, Japan\\
$^{2}$CREST, Japan Science and Technology (JST) Agency, 1-9-9 Yaesu, Chuo-ku, Tokyo 103-0028, Japan\\
$^{3}$Department of Physics, $\acute{E}$cole Polytechnique, 91128 Palaiseau Cedex, France}

\begin{abstract}
The progress in quantum operations of continuous-variable (CV) schemes can be reduced to that in CV quantum teleportation. The fidelity of quantum teleportation of an optical setup is limited by the finite degree of quantum correlation which can be prepared with a pair of finitely squeezed states. Reports of improvement of squeezing level have appeared recently, and we adopted the improved methods in our experimental system of quantum teleportation. As a result, we teleported a coherent state with a fidelity $F=0.83\pm 0.01$, which is better than any other figures reported to date. In this paper, we introduce a measure $n_{s}$, the number of teleportations expected to be carried out sequentially. Our result corresponds to $n_{s}$=5.0$\pm$0.4. It suggests that our improvement would enable us to proceed toward more advanced quantum operations involving multi-step quantum operations.
\end{abstract}

\maketitle

Quantum teleportation is a quantum operation in which one can send an arbitrary quantum state to another party by making use of the quantum correlation called {\it quantum entanglement} and classical channels between the sender and the receiver. More than that, progress in the operations of quantum states in continuous-variable (CV) schemes can be reduced to those in CV quantum teleportation \cite{Vaidman}, which can be regarded as one of the most fundamental quantum operations \cite{GottesmanChuang99,Takei,sequential,yonezawa}. Measurement-based quantum computation with cluster states \cite{Menicucci06,Peter_cluster_ex.}, and a cubic phase gate \cite{cubic_phase_gate} which involves four-time sequential teleportation \cite{Furusawa07} can be taken as examples. Therefore quantum teleportation of high quality is essential to build advanced quantum operations which involve multi-step quantum operations.

The optical scheme of CV quantum teleportation was proposed by \cite{Braunstein98}, in which squeezed states are used as a source of entanglement. Squeezed states can be generated optically via parametric down-conversion \cite{Wu86}.

In order to evaluate quantum teleportation, we introduce the fidelity as $F=\left\langle \Psi_{\mathrm{in}}|\rho_{\mathrm{out}}|\Psi_{\mathrm{in}}\right\rangle$ for a coherent state input. It can be derived as
\vspace{-7mm}
\begin{center}
\begin{equation}
F=\frac{1}{1+ne^{-2r}},
\label{fidelity_formula}
\end{equation}
\end{center}
where $r$ is the squeezing parameter of an entanglement resource and $n$ is the number of sequential teleportations \cite{Suzuki06}. 

Eq.(\ref{fidelity_formula}) shows that fidelity cannot reach $F=1$ due to the finite squeezing level, or finite degree of quantum entanglement, the so-called Einstein-Podolsky-Rosen (EPR) correlation \cite{EPR}. With $r=0$, that is without entanglement, and $n=1$, the fidelity is $F=1/2$ which can be taken as the classical limit \cite{Braunstein00,Braunstein01,Hammerer05}.

We define a figure $n_{s}$ as another measure than fidelity here. It signifies how many times teleportations are expected to be achieved sequentially. $n_{s}$ is defined so that it satisfies 1/2 $\equiv$ 1/$(1+n_{s}e^{-2r_{\mathrm{eff}}})$.  Note that $F$=1/2 is the threshold of the success of teleportation. $r_{\mathrm{eff}}$ is the effective squeezing parameter defined with an experimental fidelity as $F_{\mathrm{exp}}=1/(1+e^{-2r_{\mathrm{eff}}})$. Hence, $n_{s}$ can be expressed as
\vspace{-6mm}
\begin{center}
\begin{equation}
n_{s}= \frac{1}{e^{-2r_{\mathrm{eff}}}} = \frac{F_{\mathrm{exp}}}{1-F_{\mathrm{exp}}}.
\label{sequential_times_formula}
\end{equation}
\end{center}
This figure has more explicit meaning than the fidelity when advanced quantum operations are considered. 
 
From eq.(\ref{fidelity_formula}), higher fidelity is expected with higher squeezing level. In \cite{Furusawa98}, the first report of overcoming the classical limit, the fidelity was $F=0.58$ which could not reach $n_{s}=2$. Several reports of higher fidelity followed later \cite{Bowen03a,Zhang03,Takei,sequential,yonezawa}. For instance, Takei {\it et al}. \cite{Takei} and Yonezawa {\it et al}. \cite{sequential} reported the fidelity $F=0.70$. The former demonstrated entanglement swapping and the latter reported sequential teleportation which could be expected from eq.(\ref{fidelity_formula}), which predicts a teleporter with their fidelity surpasses $n_{s}=2$. The highest fidelity $F=0.76$ has appeared in \cite{yonezawa} recently, in which teleportation of a squeezed state was demonstrated as a result of this high fidelity. This fidelity corresponds to $n_{s}=3.1$.

Meanwhile, observations of a single highly squeezed state have been reported \cite{Aoki_SQ,Suzuki06,takeno,-10dB} in which the improvements are ascribed mainly to two ideas.

One lies in the reduction of losses in optical parametric oscillator (OPO) cavities which are used for the generation of squeezed states. It can be achieved by changing the nonlinear optical medium for production of squeezed states from $\mathrm{KNbO_{3}}$ to periodically poled $\mathrm{KTiOPO_{4}}$ (PPKTP) \cite{Aoki_SQ,Suzuki06}. $\mathrm{KNbO_{3}}$ is well known for its high nonlinearity and was reported to generate $-$6dB squeezing \cite{KN-6dB}. It is, however, also known to show blue-light-induced infrared absorption (BLIIRA) \cite{BLIIRA} which causes high intracavity loss. On the contrary, PPKTP does not show such characteristics. As a result of the change of the crystal, the remarkable improvements of the squeezing level have been reported. For example, the observation of $-$7.2dB squeezing was shown in \cite{Suzuki06}. This encouraged the use of PPKTP in \cite{sequential,yonezawa} and in our experiment as well.

Second, fluctuation of a relative phase between a squeezed state and a local oscillator causes a deterioration of squeezing by contamination with antisqueezing \cite{Aoki_SQ}. In \cite{takeno}, the phase fluctuations were reduced in addition to the small loss of PPKTP to obtain a higher squeezing level that reached $-$9.0dB. The reduction of the phase fluctuation was achieved with high oscillation frequency of the feedback loops. Recently, as much as -10dB squeezing was obtained \cite{-10dB}. These authors also pointed out the importance of phase stability.

In spite of these results, such high squeezing has not been utilized in practical quantum operations, which are much larger and more difficult to control than simple experiments on the measurement of a squeezing level. Even in \cite{yonezawa}, which reported the highest fidelity to date, the effective squeezing level was a little below -6dB.

In our experiment, we tried procedures based upon the ideas described above so that the experimental setup of quantum teleportation would be with as little losses and phase fluctuations as possible. In this paper, we show a result of $F=0.83\pm 0.01$ with these trials.

From a practical point of view, it is better to use $n_{s}$ instead of fidelity here to compare our result with previous work. From eq.(\ref{sequential_times_formula}), our result is $n_{s}$=5.0$\pm$0.4. Hence, the improvement from the fidelity with $n_{s}=3.1$ (equivalent to $F$=0.76 of \cite{yonezawa}) to ours is comparable to the one from the classical limit ($n_{s}$=1) to $F$=0.76. This means that our improvement is equivalent to all progresses reported so far in terms of $n_{s}$.

\begin{figure}[h]
  \begin{center}
    \includegraphics[width=65mm,height=40mm]{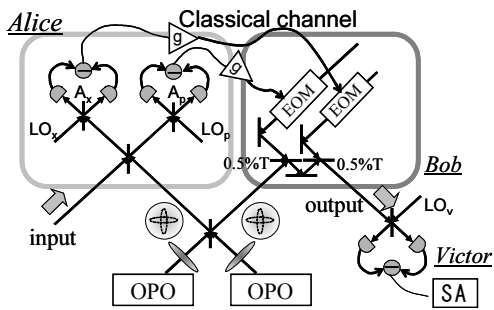}
  \end{center}
  \vspace{-5mm}\caption{Experimental setup of quantum teleportation. OPOs are optical parametric oscillators. EOMs are electro-optical modulators. Other mirrors than those labeled as 0.5$\%$T are half-beam splitters. Mirrors of 0.5$\%$ transmittance are used for displacements. LOs are local oscillators for the homodyne measurements. The output signal of Victor's homodyne measurement goes to spectrum analyzer(SA).}
  \label{fig:setup.eps}
\end{figure}
In this experiment, a quantum state of the electromagnetic field is used and can be represented by the annihilation operator $\hat{a}$. It can also be expressed with Hermite operators, $\hat{x}$ and $\hat{p}$, as $\hat{a}=\hat{x}+i\hat{p}$. Here, these two operators satisfy the commutation relationship $[\hat{x},\hat{p}]=i/2$(unit-free, with $\hbar$=1/2) and can be taken as phase and amplitude quadrature operators of the electromagnetic field.

The experimental system is shown in brief in figure \ref{fig:setup.eps}. A continuous-wave Ti:sapphire laser (Coherent MBR110, $\lambda$=860nm) is used as a light source. Smaller optical systems are expected to get more robust against fluctuation from the outside. In order to make a stable setup, it is built as compactly as possible. The optical lengths from the OPOs to the homodyne detectors are shorter than 1 m. The beam height is set as low as 2 in. also for better stability.

In order to obtain the resources of entanglement, namely, a pair of squeezed states, two OPOs are used. PPKTP crystals are used as nonlinear optical media. The OPOs are pumped by second harmonic light which is generated from a cavity which contains $\mathrm{KNbO_{3}}$ as a nonlinear medium. The pump beams are 105mW for one of the OPOs and 140mW for the other. The difference in pump power is due to individual differences of the crystals in their nonlinearities, and the parametric gains of each OPOs are set almost the same with these pump powers.

Weak coherent beams are also injected into the two OPOs. The output beams are used for locking the relative phase of the interfering beams. The output powers are set as weak as 2$\mu$W to avoid obstruction of homodyne detection. We call them the {\it probe beams}. Phase modulations of 138 and 207kHz, respectively, are applied to them with piezo-electric transdusors (PZTs) for the locking.

The lockings involve feedback loops which consist of a photodetector, a servo amplifier, and a PZT. It is important to reduce the phase delay of the signal in the loop in order to increase the oscillation frequency. The greater the oscillation frequency the feedback loop gets, the more feedback gain we can apply to reduce phase fluctuation.

In the loops, the PZTs are the most crucial component in the phase variance and they should be selected with much care. According to \cite{takeno}, the multilayered PZTs show better behavior in the feedback bandwidth than single-layer ones, and the PZTs offered by Thorlabs (AE0203D04) are shown. In fact, that type also attains the best performance in our experiment among all types we have tried. Other than that, we use small and light mirrors on the PZTs to make their response faster. The mirrors have a 1/2 in. diameter and 1.5 mm thickness, while other mirrors, which are not mounted on PZTs, have a 1 in. diameter and 6.35 mm thickness.

After preparation of the squeezed states, they are combined at a half beam splitter (HBS) with $\pi/2$ relative phase to make entangled EPR beams. They are sent to a sender Alice and a receiver Bob, respectively. Alice combines her EPR state and the input state with a HBS. This transformation can be expressed as $\hat{x_{u}}=(\hat{x}_{in}-\hat{x}_{A})/\sqrt{2}$ and $\hat{p}_{u}=(\hat{p}_{in}+\hat{p}_{A})/\sqrt{2}$. Then, she measures $\hat{x}_{u}$ and $\hat{p}_{v}$ by homodyne measurements. Bob receives the results from Alice through classical channels to displace his state in the phase space and obtains the output state. Finally, a verifier Victor evaluates the output state with homodyne detection and sends the obtained signal to a spectrum analyzer which gives the variance.

In this experiment, 1.25MHz sidebands are assigned as carriers of the quantum state. The bandwidth of the OPOs is about 10MHz. Even though the squeezing level of sidebands of higher frequency gets smaller, it is negligible around 1.25MHz (about 0.2dB from the calculation). Moreover, the Ti:sapphire beam has a bandwidth of tens of kilohertz, much smaller than MHz, and no harmonics of phase modulations applied on the probe beams are seen there. Otherwise, additional signals may contaminate the results of measurements. The method of dealing with sidebands in quantum teleportation is described in \cite{yonezawa} and \cite{broadband_teleportation}.

 Therefore, the displacement can be done with phase-modulated beams with electro-optical modulators (EOMs). These beams are combined at a beam splitter of 0.5$\%$ transmittance with Bob's EPR state. One of the beams is for the displacement of $x$ and the other for $p$. This transformation is expressed as $\hat{x}_{B}\to \hat{x}_{B}+\sqrt{2}g_{x}\hat{x}_{u}$ and $\hat{p}_{B}\to \hat{p}_{B}+\sqrt{2}g_{p}\hat{p}_{v}$ resulting in
\vspace{-7mm}
 \begin{center}
\begin{align}
\hat{x}_{out}&= \hat{x}_{in} - (\hat{x}_{A}-\hat{x}_{B}) = \hat{x}_{in} - \sqrt{2}e^{-r}\hat{x}_{vac1}\label{displacement_x}\\
\hat{p}_{out}&= \hat{p}_{in} + (\hat{p}_{A}+\hat{p}_{B}) = \hat{p}_{in} + \sqrt{2}e^{-r}\hat{p}_{vac2}
\label{displacement_p}
\end{align}
\end{center}
 with $g_{x}$=$g_{p}$=1. $\hat{x}_{vac1}$ and $\hat{p}_{vac2}$ are operators of vacuum states. With infinite correlation, $r\to \infty$, the output state is identical to the input state.

It is also necessary to set the gain of the classical channels to unity, which is indispensable to send arbitrary quantum states. This step can be achieved as follows. In the beginning, a strong phase modulation is applied on the probe beam. The phase modulation signals are sent to every party and Bob's signal is canceled via the classical channel. The intensities of electric signals applied on the EOMs are adjusted by variable attenuators in order to set the gains. These attenuators can change the attenuation with a 0.1dB step. We call this step {\it cancellation}.
Note that no pump beams are injected into the OPOs in this step, therefore no entanglement ($r=0$) is involved in the cancellation.

\begin{figure}[h]
  \begin{center}
    \begin{tabular}{cc}
      \includegraphics[width=0.5\hsize,clip]{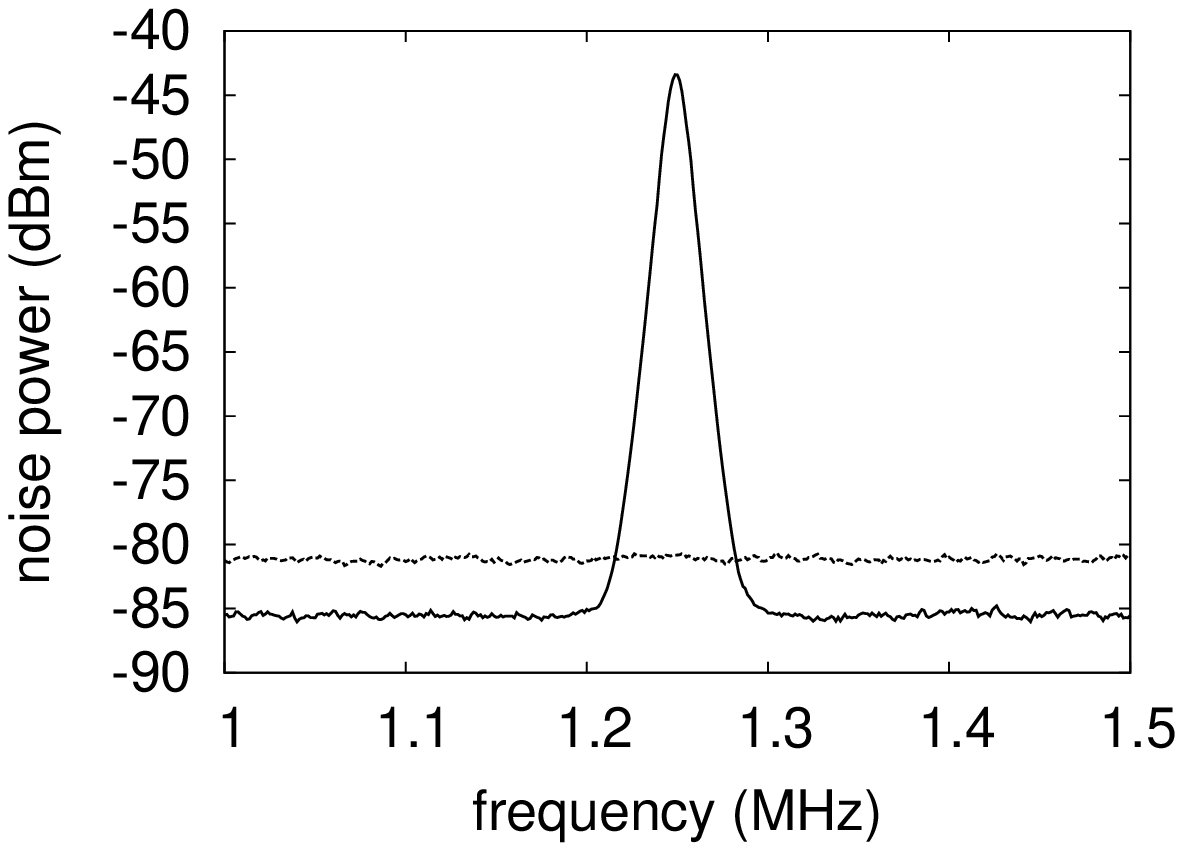}
      &
      \includegraphics[width=0.5\hsize,clip]{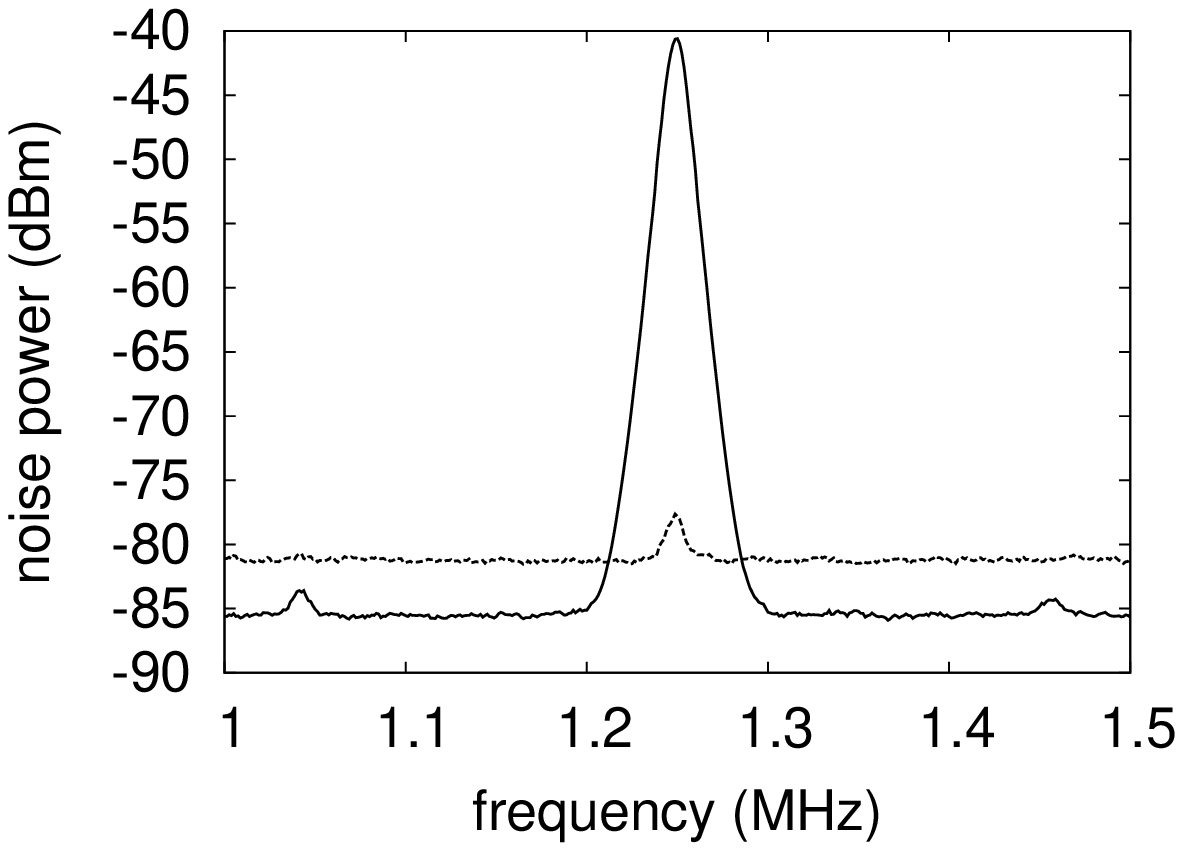} \\
     \end{tabular}
  \end{center} 
  \vspace{-5mm}
  \caption{The result of the cancellation. The left one is for $x$ quadrature and the other for $p$ quadrature. The noise level after cancellation should be 3 times as much as shot noise level as shown. The resolution bandwidth was 10kHz and all data were averaged for 20 times with 300ms of sweep time.}
    \label{cancellation}
\end{figure}

Figure \ref{cancellation} obtained from Victor's homodyne detection shows a cancellation of 37.4dB for $x$ and 37.0dB for $p$, which gives $g_{x}=1.00\pm0.01$ and $g_{p}=1.00\pm0.01$. It results in the capability of teleportation of coherent amplitude up to +37dB. Since the cancellation at frequencies other than 1.25MHz corresponds to ``classical" teleportation for a vacuum input, the noise floor of figure \ref{cancellation} increases by 4.77dB (three units of vacuum). It can be obtained with $(\hat{x}_{in},\hat{p}_{in})=(\hat{x}_{vac},\hat{p}_{vac})$ and $r$=0 in eqs.(\ref{displacement_x}) and (\ref{displacement_p}).

\vspace{2mm}

The final step is the measurement of the output state. As for fidelity, Eq.(\ref{fidelity_formula}) is used as the definition and is defined for a coherent state input. A coherent state is a gaussian state which can be fully described with its amplitude and the variances of its quadratures.

As can be seen from eqs.(\ref{displacement_x}) and (\ref{displacement_p}), the variance of an output state would be larger than that of an input state due to the finite degree of EPR correlation. With $g=1$, the complex amplitude of an output state is the same as that of an input state. Therefore, in quantum teleportation of a coherent state, only the variances of the quadratures of the output state, $\sigma^{out}_{x}$ and $\sigma^{out}_{p}$, should be obtained to evaluate quantum teleportation. The fidelity can be expressed as
\begin{center}
\begin{equation}
F=\frac{2}{\sqrt{(1+\sigma^{out}_{x})(1+\sigma^{out}_{p})}}_{,}
\label{variance_vs_fidelity}
\end{equation}
\end{center}
where $\sigma^{out}_{x}$ and $\sigma^{out}_{p}$ are normalized using a coherent state \cite{Takei,sequential}.

Hence, it is enough in the experiment to teleport one of the coherent states if $g_{x}=g_{p}=1$ is assured. This condition is warranted by the step of cancellation.

\begin{figure}[h]
  \begin{center}
    \includegraphics[width=70mm,height=50mm]{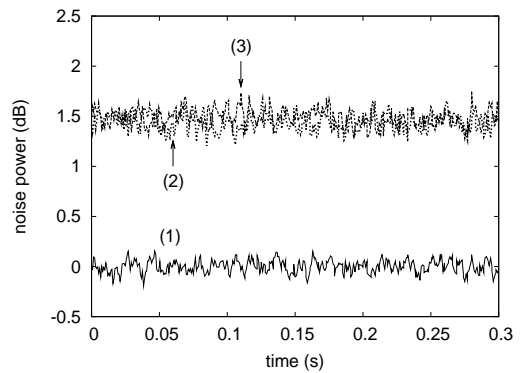}
  \end{center}
  \vspace{-5mm}
  \caption{The result of the measurement of the output state with $g$=1. (1) is shot noise level, (2)(solid line) is the variance of $x$ quadrature of the output state and (3)(dotted) is of $p$ quadrature. The measurement frequency was 1.25 MHz, resolution bandwidth was 10kHz and video bandwidth was 100Hz with 300ms of sweep time. (1) was averaged for 100 times and (2) and (3) was for 50 times.}
  \label{result_with_g=1}
\end{figure}

We choose a vacuum state as the input state, which is a coherent state with its amplitude equal to zero. The result of the teleportation is shown in figure \ref{result_with_g=1}. The measured variances of the quadratures are $\sigma^{out}_{x}$ = +1.44$\pm$ 0.11dB and $\sigma^{out}_{p}$ = +1.49$\pm$ 0.11dB, respectively, which give a fidelity $F$ = 0.83 $\pm$ 0.01 from eq.(\ref{variance_vs_fidelity}). 

We make it clear again that the precision of the unit gain of the classical channels is indispensible for teleporting arbitrary states. It should also be mentioned here that, when a vacuum state is teleported, the observed variances changes with a small change of the gain of the classical channels. For example, if $g_{x}$ and $g_{p}$ had dropped by as little as 0.01, the fidelity would have increased to 0.84. This is a fake increase and it is found that the teleporter must give a precise gain also for an accurate evaluation of fidelity.

\begin{figure}[t]
  \begin{center}
    \includegraphics[width=70mm,height=50mm]{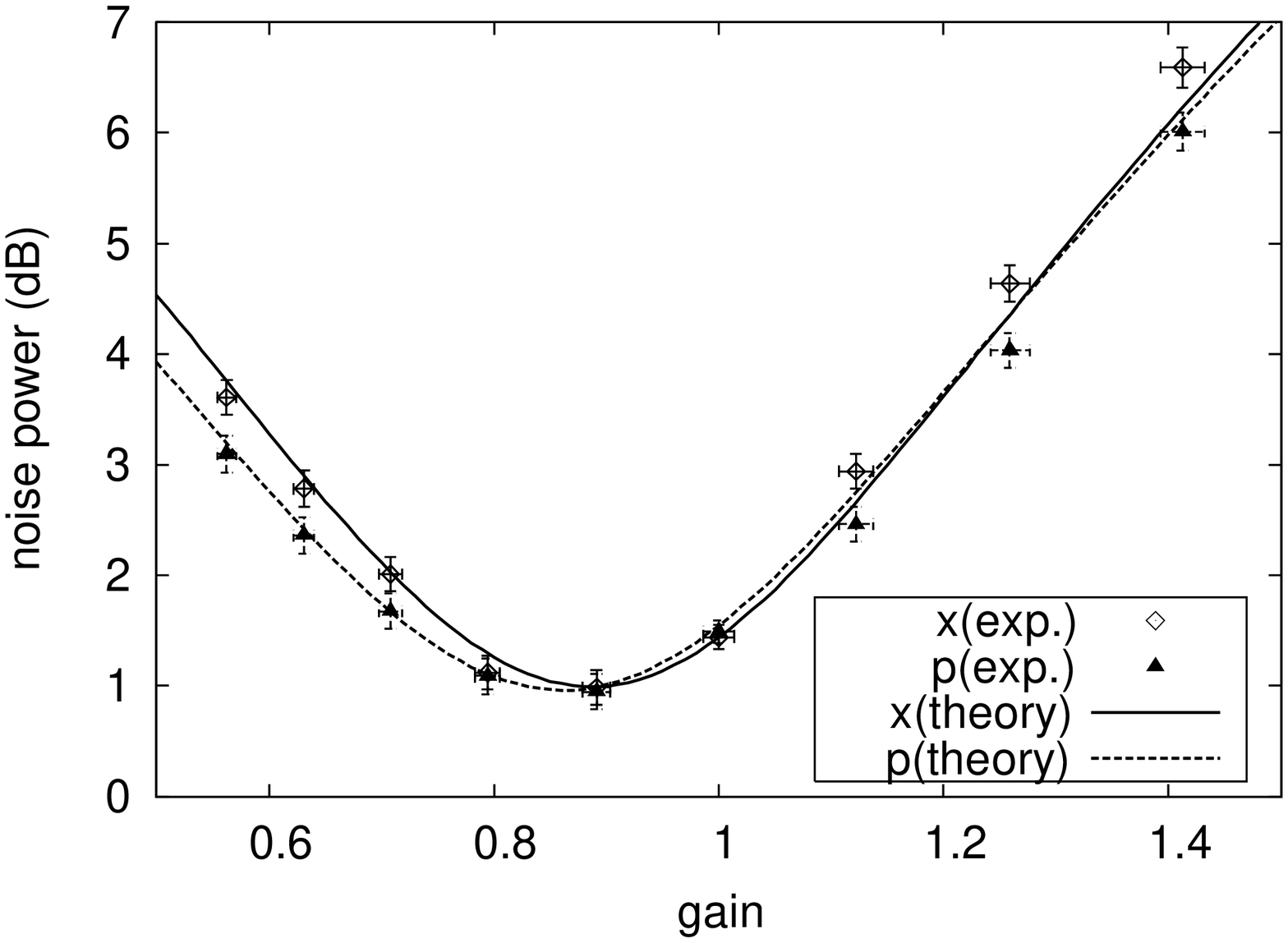}
  \end{center}
  \vspace{-5mm}
  \caption{The dependence of $\sigma_{x}$ and $\sigma_{p}$ of the output state on the gains of the classical channel, $g_{x}$ and $g_{p}$. The theoretical curves shows how output states changes when the gains are tuned. This figure shows the good agreement between theoretical values and experimental results, which garantees the result of figure \ref{result_with_g=1} is surely at $g$=1.}
  \label{fidelity_vs_gain}
\end{figure}

Although the result of the cancellation satisfies this requirement, we verify again that the gain was surely set to $g$=1 to eliminate any possibility of drifts of the gain after the cancellation. For the verification, teleportations with various gains other than $g_{x}=g_{p}$=1 were also carried out to obtain the result of figure \ref{fidelity_vs_gain}. It shows such a good agreement between the experimental results and theoretical values that it also proves that the results were obtained with unit gain. Therefore it can be concluded that we have obtained the fidelity $F=0.83\pm 0.01$.

The most dominant factor of our improvement is attributed to the reduction of the phase fluctuation, which can be explained by two changes already described.

First, the selection of the locking PZTs made the locking more rigid. The oscillation frequencies of the feedback loops were more than 10kHz, which were comparable to the results in \cite{takeno}.

Second, the experimental system was built compactly to be stronger against disturbances from the outside. We also used the least number of mirrors to avoid unexpected imperfections such as optical losses and distortions of gaussian modes.

How much each change contributed to the improvement cannot be estimated easily, but phase fluctuation of each locking was found to be suppressed to around 1$^{\circ}$, which were as good as the result achieved in \cite{takeno}.

The squeezing and antisqueezing levels of the two OPOs were also measured to be $-$7.1dB/$+$12.1dB and $-$6.8dB/$+$12.8dB. These values are the observed correlations of squeezed states split with the half beam splitter which are sent to Alice and Bob, respectively. Hence, the results include the effects of losses and phase fluctuations. These values agree well with the effective squeezing level $-$7.0dB obtained from eq.(\ref{fidelity_formula}).

Even with these improvements, these squeezing levels are apparently worse than the $-$9.0dB in \cite{takeno}. This is due to the larger total losses (about 11 $\%$) than in the measurement of a squeezed state (7 $\%$ in \cite{takeno}). The greater number of partial transmittance mirrors and worse visibilities are responsible. In addition, more number of lockings are needed, which makes the effect of anti-squeezing worse. Therefore, the parametric gains were set smaller ($G_{+}=$ 9.0 and 11.2) than in \cite{takeno} ($G_{+}=$ 18.7) and the calculation predicts that more parametric gain would give almost the same fidelity in our experiment. It can be calculated from the formulas in \cite{Aoki_SQ} that the squeezing levels should be $-$7.6 and $-$7.8dB with 11$\%$ of total loss and $\tilde{\theta}=1^{\circ}$ for each phase fluctuation. The difference from our result corresponds to merely 2 $\%$ loss.

The result is consistent with all the theoretical analysis described above. It is predicted that the quality of quantum teleportation can be improved even more through solutions of the problems discussed above. It can be achieved, for example, with more rigid locking with detectors of higher signal-to-noise ratio and treatment of all the cavity modes with more precision for better efficiency of homodyne detection.

\vspace{3mm}
In conclusion, we succeeded in obtaining the fidelity $F=0.83\pm0.01$ with a coherent state as an input state. This achievement especially owes to the stabilization of the experimental system, which is based on the ideas of \cite{takeno}. The figure $n_{s}$, the number of how many times teleportations can be carried out sequentially, was introduced as another measure than fidelity. It is more meaningful when multi-step quantum operations are implemented. $n_{s}$ of our teleporter amounts to 5.0$\pm$0.4, and in terms of $n_{s}$, our improvement is comparable to the total of all other previous improvements. It can be concluded that the improvement in this experiment is promising for progress in CV protocols toward more advanced quantum operations in the near future.

\vspace{3mm}
The authors thank N. Lee for the contributions of the improvements in the phase-locking systems. This work was partly supported by SCF and GIA commissioned by the MEXT of Japan.

\end{document}